\documentstyle[12pt]{article}
\def \d{{\rm d}}

\def \beq{\begin{equation}}
\def \eeq{\end{equation}}
\def \ln{{\rm ln}}

\def \ab2{\alpha\beta^2}

\begin{document}

\title{Dynamic fluctuations in a Short-Range Spin Glass model}
\author{Paola Ranieri}
\maketitle

\begin{center}
Dipartimento di Fisica, Universit\`a di Roma
  {\em La Sapienza}, \vspace{.1in}\\
P. Aldo Moro 2,  00185 Roma,  Italy

\begin{abstract}
We study the dynamic fluctuations of the soft-spin version of the 
Edwards-Anderson model in the critical region for $T\rightarrow T_{c}^{+}$.
First we solve the infinite-range limit of the model using the random matrix 
method. We define the static and 
dynamic 2-point and 4-point correlation functions at the order $O(1/N)$ and  
we verify that the static limit obtained from the dynamic expressions 
is correct. 
In a second part we use the functional integral formalism to
define an effective short-range Lagrangian $L$ for the fields
$\delta Q^{\alpha\beta}_{i}(t_{1},t_{2})$ up to the cubic order in the series
expansion around the dynamic Mean-Field value 
$\overline{{Q}^{\alpha\beta}}(t_{1},t_{2})$.
We find the more general expression for the
time depending non-local fluctuations, the
propagators
$[\langle\delta Q^{\alpha\beta}_{i}(t_{1},t_{2})
\delta Q^{\alpha\beta}_{j}(t_{3},t_{4})\rangle_{\xi}]_{J}$, 
 in the quadratic approximation. 
Finally we compare
the long-range limit of the correlations, derived in this formalism, with the 
correlations of the infinite-range model studied with the previous approach 
(random matrices).
\end{abstract}

\end{center}
%\hspace{.1in} PACS Numbers 7510N
\newpage

\hyphenation{dia-go-na-li-za-tion}
%INTRODUCTION
\vspace{3truecm}
\section{Introduction}
The static and dynamic Mean Field (MF) theory  of Spin Glasses (SG) systems
for $T\geq T_{c}$ is well defined and understood.
This theory has been studied through different approaches all consistent
among themselves.
Many important results concerning
the equilibrium static properties of SG have been derived using the replica
method \cite{m.p.v.}.
Sompolinsky and Zippelius \cite{s.z.1}, \cite{s.z.2}, \cite{s.}
studied a soft spin version of the Edwards-Anderson model
with the dynamic formalism, avoiding the replica trick.
They defined a Langevin dynamics on the system and
analysed the infinite range limit where the MF solution is exact.
The static limit derived from the dynamic expressions is in agreement with 
the static prevision obtained with replica. Moreover, 
dynamic characteristics of the model have been well defined.

Unfortunately
the behaviour of short-range SG system is not clear yet. There are
different analytic works and simulations about the Ising and Heisemberg
SG in finite dimensions, \cite{pa.}, \cite{og.}, \cite{y.}, \cite{em},
that do not agree with each other.\\
On the other hand, in order to study
the corrections to the MF behaviour of the Green functions
one can use the Renormalization Group theory and the $\epsilon$-expansion.
Chen, Lubensky \cite{h.l.c.}, \cite{c.l.} and Green \cite{gr.} studied a
static Ising model in
$d=6-\epsilon$ dimensions (6 is the upper critical dimension)
with the Replica method and they found the corrections to the second order
in $\epsilon$ for the critical exponents.

In this work,
we want investigate the behaviour of the dynamic fluctuations
of the short range SG in the
critical region for $T\rightarrow T^{+}$.
We study the soft-spin version of the Edwards-Anderson model that evolves
through Langevin dynamics.
We adopt the same procedure that Sompolinsky and Zippelius used in \cite{s.z.3},
and   
we manage to explicit write the time-dependent propagators
$[\langle\delta Q^{\alpha\beta}_{i}(t_{1},t_{2})
\delta Q^{\alpha\beta}_{j}(t_{3},t_{4})\rangle_{\xi}]_{J}$ for any value of
$t_{1},t_{2},t_{3},t_{4}$ in the critical region, while in \cite{s.z.3} they
were defined only in the limit of two complete time separation. 
They are the elementary building blocks in a renormalization group calculation 
expanded in $\epsilon=6-d$ and can be used for future studies of the dynamic 
effects of higher-order terms (the cubic interactions).
For evaluating these propagator we define an effective Lagrangian of the fields 
$\delta Q_{i}^{\alpha\beta}(t,t')$ (which represent the fluctuations around the 
MF order parameter value $\overline{{Q}_{i}^{\alpha\beta}}(t,t')$) through the 
functional integral formulation of the dynamics.
In order to have a comparative term order by order in perturbation theory, 
we solve the infinite-range limit and the O(1/N) corrections  of this dynamic
model, by using the 
distribution of the eigenvalues of the random interaction matrix. 
We verify that the expressions derived with the two different and independent
methods are consistent each other.
 
The aim of this work is to pursue the study of the quadratic 
fluctuations of the soft-spin model  
in the general case, without having recourse to the Glauber model (hard-spin
limit) \cite{z.}. We define the quadratic fluctuations 
%$G^{\alpha\beta\gamma\delta}(i,j;t_{1},t_{2},t_{3},t_{4})$
as a perturbative 
series in $g$ (the coupling constant of the fourth order term of the soft-spin
Lagrangian) which we succeed in resuming and therefore in obtaining a 
$g$ independent expression that could be directly used for further 
diagrammatic expansions of the theory. 
The results are qualitatively 
different from those obtained by Zippelius \cite{z.} and this may be related 
to the approximations done in going to the hard case.  
   
This paper is organized as follows:
in section 2 we define the theory; in section 3 we find the form
of the quadratic corrections to the MF limit (order $O(1/N)$, where $N$ is the
sites number) using the diagonalization of the interaction random matrix;
finally in section 4 we find the general propagator for an effective short
range Lagrangian of the field $\delta Q^{\alpha\beta}(t,t')$
(fluctuations around MF limit); in the conclusion we present the possible 
development of this work.

%FIRST SECTION
\section {Definition of the model}
Let us consider the soft spin version of the Edwards-Anderson model given
by the Hamiltonian
\begin{equation}
\beta{\cal H}=-\beta\sum_{\langle ij\rangle}\beta J_{ij}s_{i}s_{j}
           +\frac{1}{2}r_{0}\sum_{i}s_{i}^{2}+\frac{1}{4!}g
             \sum_{i}s_{i}^{4} \label{H1}\: ,
\end{equation}
where the sum $\langle ij\rangle$ is over $z$ nearest-neighbor sites
and the couplings $J_{ij}$ are quenched random variables  with the
following distribution:
\begin{equation}
      P(J_{ij})=\frac{1}{(2\pi z)^{1/2}}\exp{(-z\:J_{ij}^{2}/2)}\: .
\end{equation}
A purely relaxational dynamics can be described by the Langevin equation:
\begin{equation}
\Gamma_{0}^{-1}\frac{\partial s_{i}(t)}{\partial t}=-
\frac{\partial (\beta {\cal H})}{\partial s_{i} (t)}+\xi_{i}(t)\: .
\end{equation}
$\xi_{i}(t)$ is a Gaussian and white noise with zero mean and variance

$$\langle\xi_{i}(t)\xi_{j}(t')\rangle=\frac{2}{\Gamma_{0}}\delta_{ij}
\delta(t-t'),$$
which ensures that locally the fluctuation-dissipation theorem holds.

In the MF theory the physical quantities of interest are the
average local (at the same point) correlation function
\begin{equation}
C(t-t')=[\langle s_{i}(t) s_{i}(t')\rangle_{\xi}]_{J}\:,
\end{equation}
and the average local response function
\begin{equation}
G(t-t')=\frac{\Gamma_{0}}{2}[\langle s_{i}(t) \xi_{i}(t')\rangle_{\xi}]_{J}\: ,
\end{equation}
where the angular brackets refer to averages over the noise $\xi_{i}$
and square brackets over quenched disorder $J_{ij}$.

Beyond the MF approximation we evaluate the non local fluctuations
that are non vanishing
\begin{eqnarray}
&&\hspace{-1truecm}G^{\alpha\beta\gamma\delta}(i-j;t_{1},t_{2},t_{3},t_{4})=\nonumber\\
&&[\langle\phi_{i}^{\alpha}(t_{1})\phi_{i}^{\beta}(t_{2})\phi_{j}^
{\gamma}(t_{3})\phi_{j}^{\delta}(t_{4})\rangle_{\xi}]_{J}-
[\langle\phi_{i}^{\alpha}(t_{1})\phi_{i}^{\beta}(t_{2})\rangle_{\xi}]_{J}
[\langle\phi_{j}^{\gamma}(t_{3})\phi_{j}^{\delta}(t_{4})\rangle_{\xi}]_{J}\: ,
\end{eqnarray}
where $\alpha,\beta, \gamma, \delta$ can take the values $1$ or $ 2$ being
$\phi_{i}^{1}=\xi_{i}$ and $\phi_{i}^{2}=s_{i}$.
We shall see that these quantities represent the propagators of our theory.
We will focus on the properties of the small frequency and small momentum  
of the propagators that have a critical slowing down near $T_{c}$.
  
In this formalism the time-depending spin-glass susceptibility is  
\begin{equation}
\chi_{SG}(i-j;t_{1}-t_{3},t_{4}-t_{2})=\frac{1}{N}
[\langle s_{i}(t_{1})\xi_{j}(t_{3})\rangle_{\xi}\langle s_{j}(t_{4})
\xi_{i}(t_{2})\rangle_{\xi}]_{J}=G_{0}^{1221}(i-j;t_{1}-t_{3},t_{4}-t_{2})\:,
\label{7.1}
\end{equation}
where the $G_{0}^{\alpha\beta\gamma\delta}$ functions represent the fluctuations
in the limit $|t_{1}-t_{2}|\rightarrow\infty$ and 
$|t_{3}-t_{4}|\rightarrow\infty$. 

%SECOND SECTION
\section{Mean Field limit and quadratic fluctuations by diagonalization
of the interaction $\hspace{2cm}$ matrix}
The theory defined with the Hamiltonian (\ref{H1}) where the
$J_{ij}$ are quenched random interactions can be solved by using the general
properties of random matrices.
It is convenient to apply this approach to an infinite range model
($z=N$ and $N\rightarrow\infty$), where the non local propagators
$G^{\alpha\beta\gamma\delta}(i,j;t_{1},t_{2},t_{3},t_{4})$ represent
the corrections at order $O(1/N)$ to the MF correlations.
By an appropriated base change,
from the unidimensional spin variables
$s_{i}$ to the eigenvectors $\eta^{\alpha}$ of the $J_{ij}$ matrix,
we will manage to decouple the interaction.

We define $\eta^{\alpha}$:
$$ \sum_{j}J_{ij}\eta^{\alpha}_{j}=
\lambda^{\alpha}\eta_{j}^{\alpha},$$
where $\alpha=1......N$ and
$\lambda^{\alpha}$ is the $\alpha$-th eigenvalue.  
The properties of the eigenvalues and the eigenvectors of the random matrix
$J_{ij}$ are the following (see \cite{me.}):\\
the shape of the eigenvalues density $\sigma(\lambda)$, for symmetric
matrices with random and statistically independent
elements, such as $J_{ij}$, in the limit
$N\rightarrow\infty$ ($N$ dimension of the matrix), is:
\begin{equation}
     \sigma(\lambda)=\left\{ \begin{array}{ll}
                  \frac{1}{2\pi}(4-\lambda^{2})^{1/2}& \ \ |\lambda|<2 \\
                  0& \ \ |\lambda|>2
                \end{array}
              \right.\label{a13};
          \end{equation}
the eigenvectors are statistical variables which have the components independent
and following a gaussian distribution defined by the moments:
\begin{equation}
           \overline{\eta_{i}^{\alpha}}=0,\label{eig}
\end{equation}
\begin{equation}
              \overline{(\eta_{i}^{\alpha})^{2K}}=\frac{(2K-1)!!}{N^{K}};
\end{equation} 
they are found to be orthogonal and we can choose them to be normalized:
\begin{eqnarray}
  \sum_{i=1}^{N}\eta_{i}^{\alpha}\eta_{i}^{\beta}&=&\delta_{\alpha\beta}\;,\\
 \sum_{\alpha=1}^{N}(\eta_{i}^{\alpha})\eta_{j}^{\alpha}&=&\delta_{ij}\;;
\label{norm}
\end{eqnarray}
finally, the eigenvectors are uncorrelated among themselves (apart from the
orthogonality constraint) and they are not correlated to the eigenvalues.

If we write $s_{i}=\sum_{\alpha} Y^{\alpha}\eta_{i}^{\alpha}$
in the base of the eigenvectors we obtain:
\begin{equation}
 \beta {\cal H}_{Y}=\frac{1}{2}\sum_{\alpha}(-\beta \lambda^{\alpha}+r_{0})
                                                         (Y^{\alpha})^{2}
  +\frac{g}{4!}\sum_{\alpha \beta \gamma \delta}Y^{\alpha}Y^{\beta}
    Y^{\gamma}Y^{\delta}\sum_{i} \eta_{i}^{\alpha}\eta_{i}^{\beta}
     \eta_{i}^{\gamma}\eta_{i}^{\delta}. \label{hy}
\end{equation}
We can evaluate the Green functions for the component
$Y^{\alpha}$ and, by using (\ref{a13}) and (\ref{eig})-(\ref{norm}), go back 
to the correlation functions of $s_{i}$ field.

We can start evaluating the correlation functions in the static theory.
In the time-independent limit for the Hamiltonian (\ref{hy}) we may
consider the non-linear interaction as
a perturbation and make a series expansion in the coupling constant $g$.
We can show that only the diagrams considered in the Hartree-Fock
approximation give relevant contributions to the free theory, in
the thermodynamic limit. 
This is due to the relations (\ref{eig})-(\ref{norm}) satisfied by the
eigenvector $\eta_{i}$.
In fact, in the MF limit the relevant contribution to the interaction term in
the (\ref{hy}) is $\sum_{i}(\eta^{\alpha})^{2}(\eta^{\beta})^{2}\propto
1/N$, and in this way one can see easily that   
the eigenvector index ($\alpha$ for $Y^{\alpha}$) plays the same
role of the component index in the theories of vector fields,
where the Hartree-Fock approximation
has been demonstrated valid when the number of the components goes to infinity.
We can thus resum all the diagrams at the next orders in $g$ and
find a renormalization of the mass term (the coefficient of the quadratic term
in (\ref{hy})) that for $T\rightarrow T_{c}^{+}$ is
\begin{equation}
m^{2}\propto{\left(\frac{T-T_{c}}{T_{c}}\right)}^{2}=
                 \left(\frac{\tau}{T_{c}}\right)^{2},\label{massa}
\end{equation}
and a renormalization of the coupling constant $g$, that in the same region is
\begin{equation}
g_{r}=(2m^{2})^{\frac{1}{2}}.
\end{equation}
The static susceptibility shows a divergent behaviour for
$T\rightarrow T_{c}^{+}$
with the critical exponent $\gamma=1$, in agreement with the results obtained
with the replica method \cite{m.p.v.}
\begin{eqnarray}
\chi_{S.G.}=\frac{1}{N}\sum_{ik}\langle s_{i}s_{k}\rangle_{\sigma(\lambda)}^{2}
&=&\int_{0}^{4}\frac{d\lambda \:
      \sqrt{(4\lambda-\lambda^2)}}{2\pi(\beta\lambda+m^2)^2}\nonumber \\
&=&\frac{2}{(2m^2)^{1/2}}\propto \frac{1}{\tau}\: .\label{a18}
\end{eqnarray}
The generic $4$-point function
\begin{equation}
[\langle s_{i}s_{k}s_{i}s_{k}\rangle_{\xi}]_{J}=[\langle(s_{i})^{2}\rangle_{\xi}
\langle (s_{k})^{2}\rangle_{\xi}]_{J}+
2[(\langle s_{i}s_{k}\rangle_{\xi})^{2})]_{J}+
[\langle s_{i}s_{k}s_{i}s_{k}\rangle_{\xi\: conn}]_{J} \label{4poi}
\end{equation}
is $O(1/N)$ order and
is regular in the critical region.
In fact, for $T\rightarrow T_{c}$ the first term is regular and
the divergence of the second term
\begin{equation}
2\times\left(\frac{2}{(2m^{2})^{1/2}}\right),
\end{equation}
is compensated by that of the last one
\begin{equation}
-\left((2m^{2})^{1/2}\right)\times\left(\frac{2}{(2m^{2})^{1/2}}\right)
\times\left(\frac{2}{(2m^{2})^{1/2}}\right)=- \frac{4}{(2m^{2})^{1/2}}.
\end{equation}
Also in the replica approach one can verify that the 4-point function 
(\ref{4poi}) is not singular for $T\rightarrow T_{c}$.

In the dynamic case we have to solve the Langevin equation
for the time-dependent component $Y^{n}(t)$ (we put $\Gamma_{0}=1$):
\begin{equation}
   \dot {Y}^{n}=(\beta \lambda^{n}-r_{0})Y^{n}-g\sum_{\alpha,\beta,\gamma}
     Y^{\alpha}Y^{\beta}Y^{\gamma}\sum_{i}\eta_{i}^{\alpha}\eta_{i}^{\beta}
       \eta_{i}^{\gamma}\eta_{i}^{n}+\xi^{n}\;,
\end{equation}
If we define $G^{n}(t-t')=1/2\langle Y^{n}(t)\xi^{n}(t')\rangle$ and 
$C^{n}(t-t')=\langle Y^{n}(t)Y^{n}(t')\rangle$ we find 
the formal solution 
\begin{eqnarray}
  &&\displaystyle{Y^{n}(t)=\int_{0}^{t}G^{n}(t-t')\xi^{n}(t')dt'+} \nonumber\\
&&\displaystyle{-g\int_{0}^{t}dt'G^{n}(t-t')
  \sum_{\alpha,\beta,\gamma}Y^{\alpha}(t')Y^{\beta}(t')Y^{\gamma}
  (t')\sum_{i}\eta_{i}^{\alpha}\eta_{i}^{\beta}\eta_{i}^{\gamma}
   \eta_{i}^{n}.} \label{Yg}
\end{eqnarray}
This is a self-consistent equation, which can be solved with an iterative
procedure. At finite order in $g$, $Y^{n}$ is a polynomial in
the variables $\xi^{n}(t)$.
We can obtain the dynamic physical quantities averaging $G^{n}$ and $C^{n}$ 
over the distribution of the eigenvalues (\ref{a13}) and of the eigenvectors
defined by (\ref{eig})-(\ref{norm}). We can show that, as in the static case, 
only the diagrams of the Hartree-fock type are relevant
in the correction to the Green functions. 
The correction 
terms do not change the dynamic behaviour of the $2$-point functions 
for $T=T_{c}$, because
the time dependent part of the self-energy is regular at $T=T_{c}$ \cite{s.z.2}.
After some algebraic calculation we find only a mass renormalization
effect for $C(\omega)$ and $G(\omega)$
\begin{eqnarray}
G_{r}(\omega)&=&\int\frac{d\lambda\:\sigma(\lambda)}
      {(-\beta\lambda+m^2-i\omega)}\nonumber\\ 
&=&2+\Delta T_{c}+\tau-2\sqrt{2(m^2-i\omega)}\;, \label{1g}\\
            \nonumber\\
C_{r}(\omega)&=&\int\frac{d\lambda\:\sigma(\lambda)}
      {((-\beta\lambda+m^2)^2-i\omega^2)}\nonumber\\  
&=&\frac{2\cdot 4}
{\sqrt{2(m^2-i\omega)}+\sqrt{2(m^2+i\omega)}}\;.
\label{1c} \end{eqnarray}
where $m^2$ is the renormalized static parameter (\ref{massa}), 
$\Delta T_{c}=T_{c}-T_{c}^{0}$ is the difference from the renormalized
and the bare critical temperature ($T_{c}^{0}=2$), while as usually 
$\tau=T-T_{c}$.
The relations (\ref{1g}) and (\ref{1c}) are in agreement with the critical
behaviour indicated in \cite{s.z.2} by Sompolinsky and Zippelius.  
The susceptibility, according to the definition (\ref{7.1}), results
\begin{equation}
\chi_{S.G.}(\omega_{1},\omega_{2})=\int\frac{d\lambda\:\sigma(\lambda)}
      {(-\beta\lambda+m^2-i\omega_{1})(-\beta\lambda+m^2-i\omega_{2})}. 
\end{equation} 
It is clear that $\chi_{SG}(\omega_{1},\omega_{2})$ has a finite limit for
$\omega\rightarrow 0$ when $T>T_{c}$, while at $T=T_{c}$ it shows a critical 
behaviour such as 
\begin{equation}
\chi\propto \frac{1}{\omega^{1/2}}.
\end{equation}

For the generic 4-point
function the connected terms occur.
For example, for 
$\langle \xi_{i}(t_{1})s_{i}(t_{2})s_{k}(t_{3})\xi_{k}(t_{4})\rangle$, we have
the diagram of Fig.[1], where, as usual, a line with an arrow represents
a response function(the time order follows the arrow) and one with a cross a 
correlation function.  
The diagrams that we have to sum in order to evaluate the renormalized coupling
constant are drawn in Fig.[2] and in the low frequency
limit we have:
\begin{equation}
g_{r}=\frac{1}{2}\sqrt{2}\sqrt{2(2m^{2}-i\omega)}.
\end{equation}
Therefore the total contribution to $\langle \xi_{i}s_{i} s_{k}\xi_{k}\rangle$
function, represented in Fig.[3], is the sum of the following terms:
\begin{eqnarray}
1)&\displaystyle{-\frac{1}{2}\sqrt{2}\left(\sqrt{2(2m^2-i\bar{\omega}}\right)
\frac{4}{\sqrt{2(m^2-i\omega_{1})}+\sqrt{2(m^2+i\omega_{2})}}}\times&\nonumber\\
&\displaystyle{\frac{1}{\sqrt{2(m^2+i\omega_{3})}+\sqrt{2(m^2+i\omega_{4})}}
\frac{1}{\sqrt{2(m^2-i\omega_{3})}+\sqrt{2(m^2+i\omega_{4})}}}\times&\nonumber\\
&\displaystyle{\frac{16}{\sqrt{2(m^2-i\omega_{3})}+\sqrt{2(m^2+i\omega_{3})}}},
& \label{1c1}
\end{eqnarray}
\begin{eqnarray}
2)&\displaystyle{-\frac{1}{2}\sqrt{2}\left(\sqrt{2(2m^2+i\bar{\omega}}\right)
\frac{4}{\sqrt{2(m^2-i\omega_{3})}+\sqrt{2(m^2+i\omega_{4})}}}\times&\nonumber\\
&\displaystyle{\frac{1}{\sqrt{2(m^2+i\omega_{1})}+\sqrt{2(m^2+i\omega_{2})}}
\frac{1}{\sqrt{2(m^2-i\omega_{1})}+\sqrt{2(m^2+i\omega_{2})}}}\times&\nonumber\\
&\displaystyle{\frac{16}
{\sqrt{2(m^2-i\omega_{1})}+\sqrt{2(m^2+i\omega_{1})}}},&\label{2c1}
\end{eqnarray}
\begin{eqnarray}
3)&\displaystyle{\frac{1}{2}\left(\frac{1}{2}\sqrt{2}\sqrt{2(2m^2-i\bar{\omega}
)}\right)
\left(\frac{1}{2}\sqrt{2}\sqrt{2(2m^2+i\bar{\omega})}\right)}\times&\nonumber\\
&\displaystyle{\frac{2i\sqrt{2}}{\bar{\omega}}\left(\frac{1}{\sqrt{2(m^2-i
\bar{\omega})}}
-\frac{1}{\sqrt{2(m^2+i\bar{\omega})}}\right)
\frac{4}{\sqrt{2(m^2-i\omega_{1})}+\sqrt{2(m^2+i\omega_{2})}}}\times&\nonumber\\
&\displaystyle{\frac{4}{\sqrt{2(m^2-i\omega_{3})}+\sqrt{2(m^2+i\omega_{4})}}}.&
\label{3c1}
\end{eqnarray}
In the same way we can evaluate all the 4-point functions.
The static limit derived from the dynamic functions coincides, order by order
in $g$,
with the static results. This is true also for the renormalized functions.

It is not easy to extend this formalism to the short-range model, because in
that case the eigenvalue density $\sigma(\lambda)$ is not given by the simple
expression (\ref{a13}).
Although the results of this section have been useful to us for understanding
some properties of the correlation functions in the critical region, 
to go beyond the MF approximation we have to leave this approach.
In the next section we will use the functional integral formalism for a dynamic
SG model in finite dimensions.
The results obtained in this section will be used as
comparative terms for the long-range limit of the short-range correlation
functions.

%THIRD SECTION
\section{Fluctuations in short range spin glasses.$\hspace{3cm}$
Functional integral formulation.}
To solve a theory with quenched parameters we can use the functional integral
formulation for dynamics, introduced by De Dominicis \cite{dom.pel.}.
So we consider an auxiliary field $\hat{s}_{i}(t)$, \cite{m.s.r.},
and we define a two-component vector field \\
$\phi_{i}^{\alpha}=(i\hat{s}_{i},s_{i})$. One can show \cite{s.z.2}
that in this formalism the factor $i\hat{s}_{i}(t)$ in a correlation function 
acts like 
$\frac{\partial}{\partial\beta h_{i}}$, where $h_{i}$ is the external field:
\begin{equation}
\langle s_{i}(t)i\hat{s}_{i}(t')\rangle_{L(s,\hat{s})}=
\frac{\partial\langle s_{i}(t)\rangle_{L(s,\hat{s})}}{\partial\beta h_{i}(t')}=
G(t-t')
\end{equation}
So $i\hat{s}_{i}(t)$ replaces the noise $\xi_{i}(t)$ to generate the response
functions.

After averaging over $J_{ij}$ (in this case we can avoid the replica trick
because the functional ${\cal Z}$ is normalized),
a 4-spin interaction is generated. It is convenient to decouple
this interaction by a gaussian transformation \cite{s.z.2}.
Then the theory can be defined with a generating functional in the
$Q_{i}^{\alpha\beta}(t,t')$ variables:
\begin{eqnarray}
{\cal Z}=\int \prod_{\alpha,\beta=1,2}[DQ_{i}^{\alpha\beta}(1,2)]
   \exp &\displaystyle{\left(-\int d1\:\d2\sum_{i,j}\tilde{K}^{-1}_{ij}Q_{i}^
{\alpha\beta}(1,2)  A^{\alpha\beta\gamma\delta} Q_{j}^{\gamma\delta}(1,2)
                                                       \right.}&\nonumber\\
  &+\left.\ln \displaystyle{\int [ds][d\hat{s}]\exp(L_{1}(s,\hat{s},Q_{i}^
{\alpha\beta})}\right)\;,& \label{c1}
\end{eqnarray}
where $\displaystyle{\tilde{K}^{-1}_{ij}=\frac{z}{\beta^2}K^{-1}_{ij}}$
($K_{ij}=1$ if $i,j$ are nearest neighbors and zero otherwise)
and  $A^{\alpha\beta\gamma\delta}$ is such that $A^{1122}= A^{2211} =A^{2112}=
A^{1221}=1$ and $0$ otherwise.\\
$L_{1}$ is:
\begin{equation}
L_{1}=L_{0}+\int d1 d2\:Q_{i}^{\alpha\beta}(12)\psi^{\alpha}_{i}(1)
                    \psi^{\beta}_{i}(2), \label{L1}
\end{equation}
where
the field $\psi_{i}^{\alpha}$ is defined:
\begin{equation}
\psi_{i}^{\alpha}=
\left(\begin{array}{cc}
0&1\\ 1&0
\end{array}\right) \phi_{i}^{\alpha}=(s_{i},i\hat{s}_{i})\;,
\end{equation}
and $L_{0}$ is the local part of the theory:
\begin{equation}
L_{0}=\int d1 \sum_{i}\left[i\hat{s}_{i}(1)\left(-\Gamma_{0}^{-1}\partial_{1}
s_{i}(1)
-r_{0}s_{i}(1)-\frac{1}{3!}g s_{1}^{3}(1)+\Gamma_{0}^{-1}i\hat{s}_{i}(1)\right)
\right]. \label{c2}
\end{equation}
The value of the Jacobi determinant $J$, associated to the integral 
formulation of the $\delta$ function \cite{dom.pel.} depends on the 
discretization chosen to regularize the
Langevin equation. With the following regularization:
\begin{equation}
s(t+\epsilon)-s(t)+\epsilon \left(-\frac{\partial(\beta
{\cal{H}})}{\partial s_{i}(t)}\right)=D_{i}^{\epsilon}(t)
\end{equation}
where $D_{i}^{\epsilon}(t)=\int_{t}^{t+\epsilon}\xi_{i}(t')\: dt'$, 
one has $J=1$ (and it has been omitted in all the previous formula).

In the previous expressions and in the followings of this section we shall
write
$1$ for $t_{1}$, $2$ for $t_{2}$ and so on.

The solution of the MF theory is well known \cite{s.z.2}, and it is consistent
with the results obtained in the previous section.
Let us consider the fluctuations around the saddle point value
$Q_{i}^{\alpha\beta}(1,2)=\overline{Q_{i}^{\alpha\beta}}(1,2)
+\delta Q_{i}^{\alpha\beta}(1,2)$.
As a result the model can be formulated in terms of the dynamic fluctuation
field $\delta Q_{i}^{\alpha\beta}(1,2)$:
\begin{equation}
{\cal Z}=\int \prod_{\alpha,\beta=1,2}D\{\delta Q_{i}^{\alpha\beta}\}
\exp(L(s,\hat{s},\overline{{Q}_{i}^{\alpha\beta}}+\delta
Q_{i}^{\alpha\beta})), \label{dq}
\end{equation}
where $L$ contains the quadratic and the cubic term of the series expansion
around the MF value:
\newpage
\begin{eqnarray}
L&=&-\sum_{1,2}\sum_{i,j}\tilde{K}^{-1}_{i,j}\delta Q_{i}^{\alpha\beta}(1,2)
  A^{\alpha\beta\gamma\delta}\delta Q_{j}^{\gamma\delta}(1,2)+\nonumber\\
  &+& \frac{1}{2}\sum_{1,2,3,4}\sum_{i}\delta Q_{i}^{\alpha\beta}(1,2)
  C^{\alpha\beta\gamma\delta}(1,2,3,4)\delta Q_{i}^{\gamma\delta}(3,4)+
                                             \nonumber\\
  &+&\frac{1}{3!}\sum_{1,2,3,4,5,6}\sum_{i}
C^{\alpha\beta\gamma\delta\mu\nu}(1,2,3,4,5,6)
  \delta Q_{i}^{\alpha\beta}(1,2)
  \delta Q_{i}^{\gamma\delta}(3,4)\delta Q_{i}^{\mu\nu}(5,6)\;.
\nonumber\\ \label{zipp}
 \end{eqnarray}
The higher order terms can be neglected because we are interested in the
behaviour of the Green functions near the critical temperature in the
paramagnetic phase.
The $C^{\alpha\beta\gamma\delta}(1,2,3,4)$ and
$C^{\alpha\beta\gamma\delta\mu\nu}(1,2,3,4,5,6)$ vertices are the 4-spin and
6-spin correlation functions of the one site MF theory described by the 
partition function
\begin{equation}
{\cal Z_{0}}=\int[ds_{i}][d\hat{s}_{i}] exp\left[L_{0}+\int d1 d2
\sum_{i}\overline
{{Q}^{\alpha\beta}_{i}}(1,2)\psi_{i}^{\alpha}(1)\psi_{i}^{\beta}(2)\right],
\label{z37}
\end{equation}
connected with respect to pairs.
For example for $C^{\alpha\beta\gamma\delta}(1,2,3,4)$ we have:
\begin{equation}
C^{\alpha\beta\gamma\delta}(1,2,3,4)=
\langle\psi_{i}^{\alpha}(1)\psi_{i}^{\beta}(2)\psi_{i}^{\gamma}(3)
\psi_{i}^{\delta}(4)\rangle_{MF}-\langle\psi_{i}^{\alpha}(1)
\psi_{i}^{\beta}(2)\rangle_{MF}\langle\psi_{i}^{\gamma}(3)
\psi_{i}^{\delta}(4)\rangle_{MF}\;.
\end{equation}

The form of the functions $C^{\alpha\beta\gamma\delta}$ and
$C^{\alpha\beta\gamma\delta\mu\nu}$ is crucial in the
following.
We want to study the universal behaviour of the system near the critical fixed 
point, so we are interesting in the singular part of these functions 
for $T=T_{c}$, and for $\omega\rightarrow 0$.
Thus we consider $T=T_{c}$ from the beginning in the MF theory  
described by the functional (\ref{z37}) for the soft-spin model. 
Because of we 
are not able to compute the 4-point and 6-point functions analytically 
in a closed simple form, 
a perturbative approach will be used. 
We perform an expansion in the quartic vertex
$\left[\frac{1}{3!}g (s_{i})^3 i\hat{s}_{i}\right]$ using 
the MF expressions (\ref{1g}) and (\ref{1c})
for the propagators 
$[\langle s_{i}(t) i\hat{s}_{i}(t')\rangle]_{J}$ and
$[\langle s_{i}(t) s_{i}(t')\rangle]_{J}$
respectively ($[\langle i\hat{s}_{i}(t)i\hat{s}_{i}(t')\rangle]_{J}\equiv 0$).

We can demonstrate that the critical behaviour of these functions is
determined only by the zero loop contributions.
In fact the loops that we can form 
with the quartic vertex of $L_{0}$, in the $\omega$ space, are:

\vspace {3cm}
\noindent
The first is infrared converging and the latter is only
logarithmic infrared diverging.
In the evaluation of any renormalized correlation function, 
these diverging loops occur always multiplied by
correlation functions which are less singular 
(for $T\rightarrow T_{c}$) than the tree level ones.
For example, for the function 
$C^{1221}(\omega_{1},\omega_{2},\omega_{3},\omega_{4})$ we obtain:
\begin{eqnarray}
&C^{1221}(\omega_{1},\omega_{2},\omega_{3},\omega_{4}) 
=G(\omega_{4}) G(\omega_{1})\delta(\omega_{4}+\omega_{2})
\delta(\omega_{1}+\omega_{3})+&\nonumber\\
&g\cdot G(-\omega_{2})G(-\omega_{3})\left[C(\omega_{1})
G(\omega_{4})+C(\omega_{4})
G(\omega_{1})\right]\delta(\omega_{1}+\omega_{2}+\omega_{3}+\omega_{4})+&
\nonumber\\
&g^2\cdot G(\omega_{1})G(\omega_{2})G(\omega_{3})G(\omega_{4})\cdot
\log(\bar{\omega})
\delta(\omega_{1}+\omega_{2}+\omega_{3}+\omega_{4}).&
\end{eqnarray}
From the expression (\ref{1g}) and (\ref{1c}) for the response and correlation
functions, we can see that the 1-loop contribution to the connected part of
$C^{1221}$ is negligible (it is of order $\log(\omega)$ in the limit
$\omega\rightarrow 0$) with respect to
the zero loop one (that is of order $1/(\omega)^{1/2}$in the same limit).   
In the same way, we find for $C^{2221}$ that only the connected part is 
different from zero and is
\begin{equation}
g G(-\omega_{2})G(-\omega_{3})G(-\omega_{1})
G(\omega_{4})
\delta(\omega_{1}+\omega_{2}+\omega_{3}+\omega_{4}).\label{c2221}
\end{equation} 
  It is reasonable to assume that this phenomenon, which we have seen in a
perturbative expansion in $g$, holds beyond the perturbative theory,
so, in order to study the critical behaviour of the system we can neglect the
loop contributions.
In \cite{z.} these correlation functions were not considered at the critical 
point and a factorized functional expression  was proposed for them in
the hard-spin limit. The functional form that we obtain is obviously
different, but we have not yet verified  if we obtain also a different 
value for the universal physical quantities (i.e. critical exponents).

The correlations of the $\phi_{i}^{\alpha}$ fields, which we are interested in,
are related to those
$\delta Q_{i}^{\alpha\beta}$ by the relations:
\begin{equation}
\displaystyle{[\langle\phi_{i}^{\alpha}(1)\phi_{i}^{\beta}(2)\rangle_{\xi}]_{J}=
2\sum_{j}(\tilde{K}^{-1})_{ij}}
\displaystyle{\left(\overline{{Q}_{j}^{\alpha\beta}}(1,2)+
\langle\delta Q_{j}^{\alpha\beta}(1,2)\rangle_{L(Q_{\alpha\beta})}\right)},
\end{equation}
\begin{eqnarray}
&\displaystyle{[\langle\phi_{i}^{\alpha}(1)\phi_{i}^{\beta}(2)\phi_{k}^
{\gamma}(3)\phi_{k}^{\delta}(4)\rangle_{\xi}]_{J}=}
\displaystyle{4\sum_{j}(\tilde{K}^{-1})_{ij}\sum_{l}
(\tilde{K}^{-1})_{kl}}&\nonumber\\
&\displaystyle{\hspace{-1cm}\langle\left(\overline{{Q}_{j}^{\alpha\beta}}(1,2)+
\delta Q_{j}^{\alpha\beta}(1,2)\right)\left(\overline{{Q}_{l}^{\gamma\delta}}
(3,4)
+\delta Q_{l}^{\gamma\delta} (3,4)\right)}\rangle_{L(Q_{\alpha\beta})}+&
\nonumber\\
&\displaystyle{\hspace{-1cm}-2(\tilde{K}^{-1})_{ik}A^{\alpha\beta\gamma\delta}
\delta(1-3)\delta(2-4)}&.
\end{eqnarray}
In the next section, therefore, we will to evaluate the correlation functions
of the fields $\delta Q_{i}^{\alpha\beta}(1,2)$.

\subsection {The Propagators}
Let us consider the expression (\ref{zipp}) with the vanishing cubic
interaction.
The generic propagator
\begin{eqnarray}
G^{\alpha\beta\gamma\delta}(i-j;1,2,3,4)=
[\langle\phi_{i}^{\alpha}(1)\phi_{i}^{\beta}(2)\phi_{j}^{\gamma}(3)
\phi_{j}^{\delta}(4)\rangle_{\xi}]_{J}-
[\langle\phi_{i}^{\alpha}(1)\phi_{i}^{\beta}(2)\rangle_{\xi}]_{J}
[\langle\phi_{j}^{\gamma}(3)\phi_{j}^{\delta}(4)\rangle_{\xi}]_{J}&& \nonumber\\
=4\sum_{l}(\tilde{K}^{-1})_{il}\sum_{k}(\tilde{K}^{-1})_{jk}
  \langle\delta Q_{l}^{\alpha\beta}(1,2)
              \delta Q_{k}^{\gamma\delta}(3,4)\rangle_{L(Q_{\alpha\beta})}-
   2(\tilde {K}^{-1})_{ij} A^{\alpha\beta\gamma\delta}\delta(1-3)
\delta(2-4)\;,&&\nonumber\\
\end{eqnarray}
is calculated in free theory and will be used to evaluate the corrections of
terms in the loop expansion.
In the free theory $G^{\alpha\beta\gamma\delta}(i-j;1,2,3,4)$
is the solution to the following integral system:
\begin{eqnarray}
\sum_{3,4}\left(-2\:\tilde{K}^{-1}(k) A^{\alpha\beta\gamma\delta}
\delta (1-3)
  \delta (2-4)+C^{\alpha\beta\gamma\delta}(1,2,3,4)\right)\times \nonumber \\
\left(\frac{1}{4}G^{\gamma\delta\mu\nu}(k;3,4,5,6)+\frac{1}{2}\tilde{K}^{-1}(k)
A^{\gamma\delta\mu\nu}\delta (3-5)\delta (4-6)\right)=\nonumber\\
=-\delta_{\alpha,\mu}\delta_{\beta,\nu}\tilde{K}^{-2}(k)
                                \delta (1-5)\delta (2-6)\; , \label{sist}
 \end{eqnarray}
For each value of $\mu$ and $\nu$ (\ref{sist}) is an integral system
of four coupled equations in the $G^{11\mu\nu}$, $G^{21\mu\nu}$,
$G^{12\mu\nu}$, $G^{22\mu\nu}$ variables.
Obviously we consider 
$G^{\alpha\beta\mu\nu}(i,j;t_{1},t_{2},t_{3},t_{4})$ invariant
under permutation of the index $\alpha,\beta,\mu,\nu$ and of the
relatives times $t_{1},t_{2},t_{3},t_{4}$.
First we solve the system for $\mu,\nu=1,2$.
We are not able to solve the system (\ref{sist}) exactly, so
we use a recursive procedure by
considering $G^{\alpha\beta\gamma\delta}$ as a perturbative series in $g$.
For $g=0$ the coefficients $C^{\alpha\beta12}(1,2,3,4)$ are factorized in the
time-difference and the integral kernel show a complete separation of the 
internal time. This is the limit in which the propagators have been computed in
\cite{s.z.3}. 
In Fourier space we obtain:
\begin{equation}
G_{0}^{1121}(k;\omega_{1},\omega_{2},\omega_{3},\omega_{4})=0\;,\label{3g1}
\end{equation}
\begin{eqnarray}
&&\hspace{-1cm}G_{0}^{2121}(k;\omega_{1},\omega_{2},\omega_{3},\omega_{4})=
\frac{\tilde{K}^{-1}(k)G(\omega_{1})G(-\omega_{2})\delta(\omega_{1}+\omega_{4})
\delta(\omega_{2}+\omega_{3})}
{\tilde{K}^{-1}(k)-G(\omega_{1})G(-\omega_{2})},\label{3g3}
\end{eqnarray}
\begin{eqnarray}
\hspace{-1.5cm}G_{0}^{2221}(k;\omega_{1},\omega_{2},\omega_{3},\omega_{4})=
\left.\frac{1}{\tilde{K}^{-1}(k)-G(\omega_{1})G(\omega_{2})}\right[
C(\omega_{3})G(-\omega_{4})\times&&\nonumber\\
\left.\frac{\tilde{K}^{-1}G(\omega_{3})G(-\omega_{4})
\left[\delta(\omega_{1}+\omega_{3})\delta(\omega_{2}+\omega_{4})+
\delta(\omega_{1}+\omega_{4})\delta(\omega_{2}+\omega_{3})\right]}
{\tilde{K}^{-1}(k)-G(\omega_{3})G(-\omega_{4})}\right].\label{3g4}&&
\end{eqnarray}
Also at the zero order of the perturbation series in $g$ we consider the mass 
term ($m^2$) renormalized by the interaction.

At low frequency and for $T\rightarrow T_{c}$ we have\\
\noindent
$\displaystyle{\tilde{K}^{-1}(k)=\frac{1}{\beta^2} K^{-1}(k)=(4+4\Delta
T_{c}+4\tau)(1+ck^2)}$, $G(\omega)$ given by the (\ref{1g}) and $C(\omega)$ by
the (\ref{1c}) and the expression
(\ref{3g3}) and  (\ref{3g4}) become:
\begin{eqnarray}
&&\hspace{-1cm}G_{0}^{2121}(k;\omega_{1},\omega_{2},\omega_{3},\omega_{4})=
\frac{4\:\delta(\omega_{1}+\omega_{4})\delta(\omega_{2}+\omega_{3})}
{ck^2+\sqrt{2(m^2-i\omega_{1}/\Gamma_{0})}+\sqrt{2(m^2+i\omega_{2}/
\Gamma_{0})}}\;,\\
                                   \nonumber\\
&&\hspace{-1cm}G_{0}^{2221}(k;\omega_{1},\omega_{2},\omega_{3},\omega_{4})=
\frac{\delta(\omega_{1}+\omega_{3})\delta(\omega_{2}+\omega_{4})+
\delta(\omega_{1}+\omega_{3})\delta(\omega_{2}+\omega_{4})}
{4(ck^2+\sqrt{2(m^2-i\omega_{1}/\Gamma_{0})}+
\sqrt{2(m^2-i\omega_{2}/\Gamma_{0})})}\times\nonumber\\
&&\displaystyle{\hspace{1cm}\frac{4}{(ck^2+\sqrt{2(m^2-i\omega_{3}/\Gamma_{0})}+
\sqrt{2(m^2+i\omega_{4}/\Gamma_{0})})}}\times\nonumber\\
&&\hspace{1cm}\frac{4\cdot 2\cdot 2}{(\sqrt{2(m^2-i\omega_{3}/\Gamma_{0})}+
\sqrt{2(m^2+i\omega_{3}/\Gamma_{0})})}\;.
\end{eqnarray}
In the same way, if we calculate the system for $\mu,\nu=2,2$ we obtain the
propagator $G^{2222}$ that for $g=0$ results:
\begin{eqnarray}
&&\displaystyle{
G_{0}^{2222}(k;\omega_{1},\omega_{2},\omega_{3},\omega_{4})=
\frac{\delta(\omega_{1}+\omega_{3})\delta(\omega_{2}+\omega_{4})}
{4(ck^2+\sqrt{2(m^2-i\omega_{1}/\Gamma_{0})}+
\sqrt{2(m^2-i\omega_{2}/\Gamma_{0})})}}\times\nonumber\\
&&\displaystyle{\hspace{1cm}
\left[\frac{2}{(ck^2+\sqrt{2(m^2+i\omega_{3}/\Gamma_{0})}+
\sqrt{2(m^2-i\omega_{4}/\Gamma_{0})})}\right.}+\nonumber\\
&&\displaystyle{\left.\hspace{3cm}
+\frac{2}{(ck^2+\sqrt{2(m^2-i\omega_{3}/\Gamma_{0})}+
\sqrt{2(m^2+i\omega_{4}/\Gamma_{0})}}\right]}\times\nonumber\\
&&\displaystyle{
\frac{2\cdot 4\cdot 2}{(\sqrt{2(m^2-i\omega_{2}/\Gamma_{0})}+
\sqrt{2(m^2+i\omega_{2}/\Gamma_{0})})}
\frac{2\cdot 4\cdot 2}{(\sqrt{2(m^2-i\omega_{3}/\Gamma_{0})}+
\sqrt{2(m^2+i\omega_{3}/\Gamma_{0})})}}\nonumber\times\\
&&\displaystyle{\hspace{2cm}
\frac{1}
{(ck^2+\sqrt{2(m^2-i\omega_{3}/\Gamma_{0})}+
\sqrt{2(m^2-i\omega_{4}/ \Gamma_{0})})}}\;,
\end{eqnarray}
plus another term of the same form proportional to
$(\delta(\omega_{1}+\omega_{4})\delta(\omega_{2}+\omega_{3}))$.

For $\chi_{S.G.}(k,\omega)=G^{2121}(k;\omega=\omega_{1}=\omega_{2})$ 
we obtain the following scaling behaviour
for $T\rightarrow T_{c}^{+}$:
\begin{equation}
\chi_{S.G.}(k,\omega)=\xi^{2-\eta}f(k\xi,\omega\xi^{z}) \label{scaling}
\end{equation}
where $\xi$ is the correlation length and $\eta$ and $z$ are the usual critical 
exponents that, at the MF approximation, take the value of $0$ and $4$
respectively.

At the first order in $g$ we consider 
$G^{\alpha\beta\gamma\delta}=G^{\alpha\beta\gamma\delta}_{0}+
G^{\alpha\beta\gamma\delta}_{1}$.
For $G^{\alpha\beta\gamma\delta}_{0}$ we use
the former solution and we solve the system in the $G_{1}$ variables.
We apply this procedure in an iterative way and we can show that,
defining from a diagrammatic point of view
\begin{eqnarray}
&G_{0}^{1221}(i-j;\omega_{1},\omega_{2},\omega_{3},\omega_{4})=
G_{0}^{1221}(i-j;\omega_{1},\omega_{2})=\hspace{6cm}&\\
\nonumber\\
&G_{0}^{2221}(i-j;\omega_{1},\omega_{2},\omega_{3},\omega_{4})=
G_{0}^{2221}(i-j;\omega_{1},\omega_{2})=\hspace{6cm}&\\
\nonumber\\
&G_{0}^{2222}(i-j;\omega_{1},\omega_{2},\omega_{3},\omega_{4})=
G_{0}^{2222}(i-j;\omega_{1},\omega_{2})=\hspace{6cm}&
\end{eqnarray}
we obtain, for the following orders in $g$, recursive expressions that can be
resumed. For istance, the terms that occur for the function
$G^{1221}(i-j;\omega_{1},\omega_{2},\omega_{3},\omega_{4})$
can be represented with the diagrams in Fig.[4].

The expansion that we obtain can be seen as a
perturbative series in the quadratic vertex               
\begin {equation}
C_{conn.}^{2221}(1,2,3,4)\delta Q^{22}_{i}(1,2)\delta
Q^{21}_{i}(34).\label{vert.}
\end{equation}
In fact, we evaluate the free propagator $G_{0}$ considering only the 
disconnected part of the $C^{\alpha\beta\gamma\delta}(1,2,3,4,)$, and we use a
perturbative approach to take account of the connected part.
The number of the topologically different vertices of form
$$C_{conn.}^{\alpha\beta\gamma\delta}(1,2,3,4)\delta Q^{\alpha\beta}_{i}(1,2)
\delta Q^{\gamma\delta}_{i}(34)$$
are $4$ (the number of the superscripts equals one or two, and 
$C^{2222}\equiv 0$).
By an explicit computation, one can see
that the diagrams obtained with the vertex (\ref{vert.}) are the most diverging
ones. 

Indeed, for example, the contribution to the function 
$\langle\delta Q^{12}(1,2)\delta Q^{12}(3,4)\rangle$ at 
the first order in g from the vertex (\ref{vert.}) is
\begin{eqnarray}
&\langle\delta Q^{12}(1,2)\delta Q^{12}(3,4)
\delta Q^{22}(5,6)\delta Q^{21}(7,8) C^{2221}_{conn}(5,6,7,8)
\rangle\propto& \nonumber\\
&\langle\delta Q^{12}(1,2)\delta Q^{22}(5,6)\rangle
\langle\delta Q^{12}(3,4)\delta Q^{21}(7,8)\rangle 
C^{2221}_{conn}(5,6,7,8)
\stackrel{k,\omega\rightarrow 0}{\longrightarrow}&\nonumber\\
&\frac{1}{\omega^{3/2}}\cdot\frac{1}{\omega^{1/2}}\cdot g \cdot 1= 
g \cdot (1/\omega^{2})&   
\end{eqnarray}
as one can verify from (\ref{1g}) (\ref{1c}) and (\ref{c2221}),
while from the vertex

$$C_{conn.}^{2121}(5,6,7,8)\delta Q^{21}(5,6)\delta
Q^{21}_{i}(7,8)$$  one obtains a weaker singularity of order: 
\begin{eqnarray}
&\langle\delta Q^{12}(1,2)\delta Q^{12}(3,4)
\delta Q^{21}(5,6)\delta Q^{21}(7,8) C^{2121}_{conn}(5,6,7,8)
\rangle\propto &\nonumber\\
&\langle\delta Q^{12}(1,2)\delta Q^{21}(5,6)\rangle
\langle\delta Q^{12}(3,4)\delta Q^{21}(7,8)\rangle 
C^{2121}_{conn}(5,6,7,8)
\stackrel{k,\omega\rightarrow 0}{\longrightarrow}&\nonumber\\
&\frac{1}{\omega^{1/2}}\cdot\frac{1}{\omega^{1/2}}\cdot g \cdot
\frac{1}{\omega^{1/2}} = g\cdot (1/\omega^{3/2}).&\label{div}   
\end{eqnarray}
At this order in $g$ we do not have any other contribution to 
$\langle\delta Q^{12}(1,2)\delta Q^{12}(3,4)\rangle$,  
because the corrections, that one can obtain with the vertices 
$$C_{conn.}^{2111}(5,6,7,8)\delta Q^{21}(5,6)\delta
Q^{11}(7,8)$$ and 
$$C_{conn.}^{1111}(5,6,7,8)\delta Q_{i}^{11}(5,6)\delta
Q^{11}_{i}(7,8)\; ,$$
involved the propagator $\langle\delta Q^{12}\delta Q^{11}_{i}\rangle$
that are vanishing at the zero order in $g$.
  
Similarly, other functions can be calculated to verify that stronger
singularity all arise  from the vertex (\ref{vert.}).

Therefore, as usual in the case of expansion in a quadratic vertex, 
we can resum the series. Moreover       
we deal with an order
parameter depending on two times and we have to 
consider, in the Fourier space, the integral over the free internal frequencies.

To evaluate the complete propagators at $T=T_{c}$,
we can define the renormalized coupling constant
\begin{equation}
g_{r}=g-g^2 I(\bar{\omega},k)+g^3 I^{2}(\bar{\omega},k)+........ =
\frac{g}{1+g\,I(\bar{\omega},k)} \label{grk1}\;,
\end{equation}
where $I(\bar{\omega},k)$, correspondent to the loop

\vspace*{2.5cm}\noindent
is given by the integral:
\newpage
\begin{eqnarray}
&\displaystyle{I_{1}(\bar{\omega},k)=\int_{-\infty}^{\infty}\frac{d\omega}{2\pi}
\frac{1}{(ck^2+\sqrt{-2i\omega/\Gamma_{0}}+
\sqrt{-2i(\bar{\omega}-\omega)/\Gamma_{0}}}}\times&\nonumber\\
&\displaystyle{\frac{1}{(ck^2+\sqrt{2i\omega/\Gamma_{0}}+
\sqrt{-2i(\bar{\omega}-\omega)/\Gamma_{0}}}
\frac{16}{\sqrt{-2i\omega/\Gamma_{0}}+\sqrt{2i\omega/\Gamma_{0}}}}\times&
\nonumber\\
&=\displaystyle{
\frac{1}{ck^2+\sqrt{-2i\bar{\omega}/\Gamma_{0}}}\;
F_{1}\left(\frac{ck^2}{\left(i\bar{\omega}/\Gamma_{0}\right)^{1/2}}
\right)}\;.&\label{grk2}
\end{eqnarray}
$F_{1}$ is a function of the variable
$\displaystyle{\left(\frac{ck^2}{(i\bar{\omega})^{1/2}}\right)}$
that exhibits a constant limit for $\omega\rightarrow 0$ and for
$k\rightarrow 0$ .
The coupling constant in the low frequency and small moments limit is,
according to (\ref{grk1}) and to (\ref{grk2}),
\begin{equation}
g_{r}=(ck^2+\sqrt{-2i\bar{\omega}/\Gamma_{0}}) 1/F_{1}\;.\label{grk3}
\end{equation}
To evaluate the total contribution of the loop corrections we must consider the
term given by the following diagram:

\vspace*{2.5cm}\noindent
with renormalized vertices.
The value of the loop is:
\begin{eqnarray}
&\hspace{-1.3cm}\displaystyle{I_{2}(\bar{\omega},k)=\int_{-\infty}^{\infty}
\left[\frac{d\omega}{2\pi}
\frac{1}{ck^2+\sqrt{-2i\omega/\Gamma_{0}}+
\sqrt{-2i(\bar{\omega}-\omega)/\Gamma_{0}}}
\frac{1}{ck^2+\sqrt{2i\omega/\Gamma_{0}}+\sqrt{2i(\bar{\omega}-\omega)/
\Gamma_{0}}}\right.}\times&\nonumber\\
&\displaystyle{\left(\frac{1}{ck^2+\sqrt{-2i\omega/\Gamma_{0}}+
\sqrt{2i(\bar{\omega}-\omega)/\Gamma_{0}}}+
\frac{1}{ck^2+\sqrt{2i\omega/\Gamma_{0}}+
\sqrt{-2i(\bar{\omega}-\omega)/\Gamma_{0}}}\right)}\times&\nonumber\\
&\displaystyle{\left.\frac{8}{\sqrt{-2i\omega/\Gamma_{0}}
+\sqrt{2i\omega/\Gamma_{0}}}
\frac{8}{\sqrt{-2i(\bar{\omega}-\omega)/\Gamma_{0}}+
\sqrt{2i(\bar{\omega}-\omega)/\Gamma_{0}}}\right]}\times&\nonumber\\
&\displaystyle{= \frac{1}{c^2k^4+2i\bar{\omega}/\Gamma_{0}}\left(\frac{1}
{ck^2+\sqrt{2i\bar{\omega}/\Gamma_{0}}}\right)
F_{2}\left(\frac{ck^2}{(i\bar{\omega}/\Gamma_{0})^{1/2}}\right)},&
\end{eqnarray}
where $F_{2}(x)$, like $F_{1}(x)$, has a constant limit for
$\omega\rightarrow 0$ and $k\rightarrow 0$.

The connected part of the free propagator $G^{1221}$ at $T=T_{c}$ is computed
adding up all the diagrams in Fig.[4].
We have:
\begin{eqnarray}
&&G^{1221}(k;\omega_{1},\omega_{2},\omega_{3},\omega_{4})_{conn}=\left[
G_{0}^{1221}(k;\omega_{1},\omega_{2})G_{0}^{2122}(k;\omega_{3},\omega_{4})
g_{r}(k;\bar\omega)\frac{1}{F_{1}}+\right.\nonumber\\
&&\hspace{2truecm}+G_{0}^{1222}(k;\omega_{1},\omega_{2})
G_{0}^{2121}(k;\omega_{3},\omega_{4})g_{r}(k;-\bar\omega)
\frac{1}{F_{1}}+\nonumber\\
&&+G_{0}^{1221}(k;\omega_{1},\omega_{2})G_{0}^{2121}(k;\omega_{3},\omega_{4})
      g_{r}(k;\bar\omega)g_{r}(k;-\bar\omega) \frac{1}{F_{1}^2}\nonumber\\
&&\left.\hspace{2truecm}\frac{1}{c^2k^4+2i\bar{\omega}/\Gamma_{0}}\left(\frac{1}
{ck^2+\sqrt{2i\bar{\omega}/\Gamma_{0}}}\right)
F_{2}\right]\delta (\omega{1}+\omega{2}+\omega{3}+\omega{4})\nonumber\\   
\label{prop}
\end{eqnarray}
which corresponds to the diagrams in Fig.[5] with the constant $g_{r}$ given by
(\ref{grk3}) and $\bar{\omega}=\omega_{2}-\omega_{1}$.

%CONCLUSION
\section{Conclusion}
The connected term of the propagators is zero only in the limit of two 
complete time separations.
In all the other cases we must calculate the complete correlation function.
In this way we compute $G^{\alpha\beta\gamma\delta}$
for all the values of the indices $\alpha\beta\gamma\delta$ and for 
all the time distances.
It is easy to see that
the long-range limit ($k\rightarrow 0$) of the connected part of the 
expressions (\ref{prop})
coincides with the sum of the terms (\ref{1c1}) (\ref{2c1}) (\ref{3c1}) of the
third section.
On the level of the Gaussian approximation, i.e. cubic interactions are 
neglected, the connected part of the propagators does not contribute 
to the susceptibility and the dynamic scaling (\ref{scaling}) is correct.   

Therefore, we have analysed the critical behaviour of the propagators of 
the soft-spin model in the quadratic approximation and we have put the bases
for a short-range theory of SG in the renormalization group formalism. 
In fact the expressions that we have derived could be used to
evaluate the contributions of the Feynmann diagrams that occur in the loop 
expansion when the cubic term of the Lagrangian (\ref{zipp}) is considered 
non vanishing.
The expression which we have obtained in the soft-spin case are quite different 
from those obtained by Zippelius in ref. \cite{z.}. It is certainly 
interesting to understand if the value of the dynamical critical exponent is 
affected by this difference. The computation of the loops will be crucial to 
clarify this point.

\section*{Acknowledgments}
I am grateful to Giorgio Parisi for his support, essential for the realization
of this work. I also would like to thank Enzo Marinari for useful discussion 
and suggestion.

\newpage

% REFERENCES

\vspace{15cm}
%\newpage
%\supereject      

\begin{table}

\begin{tabular}{c}
CAPTIONS FOR ILLUSTRATIONS
\end{tabular}

\vspace{1cm}

\begin{tabular}{ll}

Fig. 1: & Diagrams for the correlation function
                  $[\langle\xi_{i}s_{i}s_{k}\xi_{k}\rangle]$\\&
                 in the series expansion in $g$.

\vspace{.5cm} \\

Fig. 2: & 1-particle irreducible diagrams that contribute to\\&
           the renormalization of the coupling constant.

\vspace{.5cm}\\

Fig. 3: &The renormalized function
             $[\langle \xi_{i}s_{i}s_{k}\xi_{k}\rangle]_{conn}$.

\vspace{.5cm}\\

Fig. 4: &Diagrammatic representation of the propagator\\&
$G^{1221}(i-j;\omega_{1},\omega_{2},\omega_{3},\omega_{4})$ in the series
expansion in $g$.

\vspace{.5cm}\\

Fig. 5: &The connected part of the propagator\\&
$G^{1221}(i-k;\omega_{1},\omega_{2},\omega_{3},\omega_{4})$.

\vspace{10cm}\\
\end{tabular}
\end{table}
\newpage
\end{document}